\documentclass[twocolumn,showpacs,preprintnumbers,amsmath,amssymb,APSl,prd,nofootinbib,superscriptaddress]{revtex4-1}

\usepackage{dcolumn}
\usepackage{bm}
\usepackage{ifpdf}
\usepackage{hyperref}
\usepackage{dcolumn}
\usepackage{bm}
\usepackage[spanish,english]{babel}
\usepackage{amsfonts}
\usepackage{amssymb}
\usepackage{graphicx}
\usepackage{epstopdf}
\usepackage[latin1]{inputenc}

\newcommand \be{\begin{equation}}
\newcommand \en{\end{equation}}
\newcommand \bea{\begin{eqnarray}}
\newcommand \ena{\end{eqnarray}}

\begin{document}

\title{Black hole solutions in functional extensions of Born-Infeld gravity}

\author{Cosimo Bambi}\email{bambi@fudan.edu.cn}
\affiliation{Center for Field Theory and Particle Physics and Department of
Physics, Fudan University, 220 Handan Road, 200433 Shanghai, China}
\affiliation{Theoretical Astrophysics, Eberhard-Karls Universit\"at T\"ubingen,
Auf der Morgenstelle 10, 72076 T\"ubingen, Germany}
\author{D. Rubiera-Garcia} \email{drgarcia@fc.ul.pt}
\affiliation{Instituto de Astrof\'isica e Ci\^encias do Espa\c{c}o, Universidade de Lisboa, Faculdade de Ci\^encias, Campo Grande, PT1749-016 Lisboa, Portugal}
\author{Yixu Wang}
\affiliation{Center for Field Theory and Particle Physics and Department of
Physics, Fudan University, 220 Handan Road, 200433 Shanghai, China}


\date{\today}

\begin{abstract}
We consider electrovacuum black hole spacetimes in classical extensions of Eddington-inspired Born-Infeld gravity. By rewriting Born-Infeld action as the square root of the determinant of a matrix $\hat{\Omega}$, we consider the family of models $f( \vert\hat{\Omega}\vert)$, and study black hole solutions for a power-law family of models labelled by a simple parameter. We show how the innermost structure of the corresponding black holes is modified as compared to their General Relativity counterparts, discussing in which cases a wormhole structure replaces the point-like singularity. We go forward to argue that in such cases a geodesically complete and thus non-singular spacetime is present, despite the existence of curvature divergences at the wormhole throat.
\end{abstract}

\maketitle

\section{Introduction}

Nowadays we have large experimental evidence supporting the standard $\Lambda$CDM model, which is based on Einstein's General Theory of Relativity (GR), a cold dark matter contribution and a tiny cosmological constant \cite{LCDM}. Nonetheless there are several difficulties associated with the theoretical basis of this framework. On the one hand, there are several theorems predicting the unavoidable existence of singularities deep inside black holes and in the early universe \cite{theorems}, where spacetime geometry breaks down and Physics looses predictability. On the other hand, one also faces the ad hoc character of the dark matter/energy sources since, despite the fact that many proposals to identifying their nature have been introduced along the years (see e.g. \cite{Bertone} for some reviews), there has been no direct experimental evidence at particle accelerators so far, though projects of indirect astronomical searches are ongoing \cite{Searches}. An alternative approach, instead of proposing unknown dark matter/energy components, is represented by the theoretical attempts to address these issues by enlarging the gravitational sector, which has been considered in the literature according to different perspectives. For example, high-energy modifications of GR have been studied within the theory of quantized fields in curved spacetimes \cite{quant}, from fundamental unifying theories such as string theory \cite{strings} or loop quantum gravity \cite{LQG}, or though brane-world models \cite{brane}.

It should be noted that the renormalizability of the matter fields in curved spacetimes requires a high-energy completion of the Einstein-Hilbert Lagrangian of GR involving quadratic (and higher order) terms in curvature \cite{quant}. Unfortunately, these contributions generically involve higher-order derivative equations (with the notable exception of the well known class of Lovelock gravities \cite{Lovelock}), which implies the existence of ghost-like particles, besides making it much more difficult to find exact solutions. Nonetheless, many models like the popular $f(R)$ ones, have been widely investigated with the purpose of phenomenological applications in inflation, dark energy, late-singularities and so on \cite{fRreview,fR}.

It is worth pointing out that for decades many other approaches to gravitational physics has been pursued, of which we bring forward Eddington's proposal \cite{Eddington}. It is based on the consideration of actions without the need of a metric (purely affine approach), namely

\begin{equation}
S_{Edd}=\frac{1}{\kappa^2\epsilon}\int d^4x \left[\sqrt{\vert R_{(\mu\nu)}(\Gamma) \vert}\right] \ ,
\end{equation}
where vertical bars denote a determinant and $R_{(\mu\nu)}$ is the symmetric part of the Ricci tensor, which only depends on the affine connection $\Gamma \equiv \Gamma_{\mu\nu}^{\lambda}$. However, the addition of matter to this action yields a theory whose dynamics is fully equivalent to the Einstein-Hilbert Lagrangian of GR with a cosmological constant term, as was shown in Ref.\cite{Ferraris}. To overcome this degeneracy problem of the solutions, Deser and Gibbons proposed an extension of Eddington's action using a model dubbed \emph{Born-Infeld gravity} \cite{Deser} (later studied in detail in Ref.\cite{BF}), and which on its simplest version is defined by the action

\begin{eqnarray} \label{eq:EiBI}
S_{BI}&=&\frac{1}{\kappa^2\epsilon}\int d^4x \left[\sqrt{-|g_{\mu\nu}+\epsilon R_{\mu\nu}(\Gamma)|}-\lambda \sqrt{-|g_{\mu\nu}|}\right] \nonumber \\
&+& S_m(g_{\mu\nu},\psi_m) \ .
\end{eqnarray}
This action can be recognized as the gravitational counterpart of Born-Infeld theory of electrodynamics \cite{BIem}, where Maxwell field strength tensor $F_{\mu\nu}$ is replaced by the Ricci tensor $R_{\mu\nu}$. The following definitions apply: $\epsilon$ is a small parameter with dimensions of length squared, $\kappa^2=8\pi G$ is Newton's constant, $\lambda$ is a constant which is related to an effective cosmological constant (this will be explained in detail later), $g_{\mu\nu}$ is the space-time metric and $S_m(g_{\mu\nu},\psi_m)=\int d^4 x \sqrt{-g} L_m$ is the matter sector, where the matter Lagrangian, $L_m(g_{\mu\nu},\psi_m)$, is assumed to couple only to the metric $g_{\mu\nu}$, and $\psi_m$ denotes collectively the matter fields. In much the same way that Born-Infeld action for matter removes the divergence of electron's self energy in classical electrodynamics, one could wonder whether similar mechanisms could operate on the gravitational side to remove spacetime singularities\footnote{Note in passing by that the coupling of Born-Infeld and other theories of electrodynamics to GR has been unable to provide a consistent resolution to the issue of spacetime singularities, see e.g. \cite{NED} for some literature.}. In this sense, it is worth noting that Born-Infeld gravity has been found to have many applications in astrophysics and cosmology \cite{BI-papers}.

Following the spirit of Eddington's approach, Born-Infeld gravity is formulated in the \emph{Palatini approach}, where metric and connection are regarded as independent degrees of freedom. This means that, as opposed to the standard metric approach, the connection is not constrained \emph{a priori} to be given by the Christoffel symbols of the metric [see e.g. \cite{Olmoreview} for a detailed description of this approach and \cite{Zanelli} for a pedagogical explanation]. We point out that the question of whether the underlying structure of spacetime is Riemannian or not (or, in other words, whether the geometry is completely determined by the metric degrees of freedom or non-metricity is present) is as a fundamental question as the number of spacetime dimensions or the existence of supersymmetry, but which has received comparatively little attention until very recent times \cite{Zanelli}. Interestingly, current research using the Palatini approach in several theories of gravity, like $f(R)$ \cite{or1}, with Ricci-squared corrections \cite{or2} and in higher-dimensional generalizations \cite{or3},  has explicitly shown that these theories yield second-order field equations for the metric, absence of ghost-like instabilities, and recovery of Einstein equations in vacuum. Moreover, it has been found that Born-Infeld gravity with $\epsilon <0$ yields nonsingular black hole solutions when coupled to an electromagnetic field \cite{ors}, and to bouncing cosmologies \cite{oor} [we note that Ba\~nados and Ferreira in Ref.\cite{BF} studied the corresponding solutions with $\epsilon >0$, with the result that a singularity is always present].

A natural question in this context is to what extend the singularity avoidance is attached to the particular functional form of Born-Infeld gravity or, in other words, how robust is this result. It is thus important to investigate further gravitational actions in looking for nonsingular solutions. In this sense, by noting that Born-Infeld action (\ref{eq:EiBI}) can be rewritten as

\begin{equation} \label{eq:EiBI2}
S_{BI}=\frac{1}{\kappa^2\epsilon}\int d^4x \sqrt{-g} \left[ \sqrt{ \vert \hat{\Omega} \vert } -\lambda \right] + S_m(g_{\mu\nu},\psi_m) \ ,
\end{equation}
where $g$ denotes the determinant of $g_{\mu\nu}$, the object $\hat{\Omega} \equiv \hat{g}^{-1} \hat{q}$ (here a hat denotes a matrix representation) and $q_{\mu\nu} \equiv g_{\mu\nu}+\epsilon R_{\mu\nu}$,  new models with an additional trace term were considered in \cite{Bou}, while in \cite{oor} a family of polynomial $f(\vert \hat{\Omega} \vert)$ models were studied, both with regard to bouncing cosmologies. In both cases it was found that the existence of bouncing solutions is robust against changes in the equation of state. The main aim of this work is to consider spherically symmetric black hole spacetimes with a family of power-law $f(\vert \hat{\Omega} \vert)$ models labelled by a single parameter, which recovers GR at low energies and are analytically tractable, and to investigate in which cases do wormhole structures exist. This goes beyond previous results for both the original Born-Infeld gravity case \cite{ors} and of Palatini gravity containing quadratic curvature corrections on the curvature scalar and Ricci-squared terms \cite{or2} (as we shall see, the latter arises in the low-energy expansion of the models considered here, see Eq.(\ref{eq:action-low}) below). In those cases where wormhole structures arise, we show that the corresponding spacetime is geodesically complete, which we argue makes such solutions to be regarded as non-singular. This is so despite the fact that some of the curvature scalars generically diverge on a sphere of radius the wormhole throat, but these curvature divergences do not prevent the completeness of geodesics.

This work is organized as follows: in Sec.\ref{sec:II} we introduce the action and conventions, and cast the field equations in suitable form for calculations. These field equations are explicitly written for electrovacuum solutions in Sec.\ref{sec:III} and solved for a particular family of theories in Sec.\ref{sec:IV}. The resulting solutions are characterized in detail in Sec.\ref{sec:V}, where we pay special attention to their geodesic completeness. Sec.\ref{sec:VI} contains our conclusions and some perspectives for future work.

\section{Action and field equations} \label{sec:II}

The action of our theory is defined as follows:

\be \label{eq:action}
S=\frac{1}{k^2 \epsilon} \int d^4x \sqrt{-g} [f(\vert \hat{\Omega} \vert) - \lambda] + S_m(g_{\mu\nu},\psi_m) \ ,
\en
with the definitions above. Note that the Ricci tensor, defined as $R_{\mu\nu} \equiv {R^\alpha}_{\mu\alpha\nu}$, where

\be
{R^\alpha}_{\beta\mu\nu}=\partial_{\mu}
\Gamma^{\alpha}_{\nu\beta}-\partial_{\nu}
\Gamma^{\alpha}_{\mu\beta}+\Gamma^{\alpha}_{\mu\lambda}\Gamma^{\lambda}_{\nu\beta}-\Gamma^{\alpha}_{\nu\lambda}\Gamma^{\lambda}_{\mu\beta} \ ,
\en
is the Riemann tensor, is entirely constructed out of the connection $\Gamma \equiv \Gamma^{\lambda}_{\mu\nu}$, which is a priori independent of the metric $g_{\mu\nu}$.  The action (\ref{eq:action}) represents a natural generalization of the Eddington-inspired Born-Infeld gravity (\ref{eq:EiBI}), which is obtained as a particular case of Eq.(\ref{eq:action}), $f_{BI}= \vert \hat{\Omega} \vert^{1/2}$.

To obtain the field equations we conveniently introduce a scalar field, $A$, and rewrite the action (\ref{eq:action}) as

\begin{equation}\label{eq:st-action}
S_f=\frac{1}{\kappa^2\epsilon}\int d^4x \sqrt{-g}\left[f(A)+(|\hat\Omega|-A)f_A-\lambda \right]+S_m \ ,
\end{equation}
This is indeed the action of a scalar-tensor theory, where the scalar field and the potential are given by

\be
\phi=\frac{df}{dA} \hspace{0.1cm} ; \hspace{0.1cm} V(\phi)=A(\phi)f_A - f(A) \ ,
\en
respectively. When a Lagrangian density $f(A)$ is given, one is able to construct explicitly both $\phi$ and $V(\phi)$. As described in detail in Ref.\cite{oor}, the field equations for this theory can be obtained through variation of (\ref{eq:st-action}) with respect to metric, connection and scalar field, which yields

\begin{eqnarray}\label{eq:gvar}
\phi |\hat\Omega| {q}^{\mu\nu}-\frac{\left(\phi |\hat\Omega|+V(\phi)+\lambda\right)}{2}g^{\mu\nu}&=&-\frac{\kappa^2\epsilon}{2}T^{\mu\nu} \\
\nabla_\lambda\left[\phi \sqrt{-g} |\hat\Omega| {q}^{\mu\nu}\right]&=&0\\ \label{eq:connection1}
\left(|\hat\Omega|-\frac{dV}{d\phi}\right)&=&0 \ . \label{eq:V_phi}
\end{eqnarray}
where $T^{\mu\nu}=-\frac{2}{\sqrt{-g}} \frac{\delta S_m}{\delta g_{\mu\nu}}$ is the energy-momentum tensor of the matter. To solve these equations we proceed as follows: first we note that (\ref{eq:V_phi}) simply establishes an algebraic relation between the fundamental object $\vert \hat{\Omega} \vert$ and the potential $V(\phi)$, which allows to express $\phi$ as a function of $\vert \hat{\Omega} \vert$ once the Lagrangian density $f(A)$ is given. On the other hand, from the definition $\hat\Omega={\hat g}^{-1}\hat q$ we take the determinant on both sides to obtain
 $|\hat\Omega|=|q|/|g|$, which implies $\sqrt{-g}=\sqrt{-q}\vert \Omega \vert^{-1/2}$. Using this result we can conveniently rewrite the equation for the connection (\ref{eq:connection1}) as

\begin{equation}\label{eq:connection3}
\nabla_\lambda\left[\sqrt{-t}  {t}^{\mu\nu}\right]=0 \ ,
\end{equation}
where we have introduced the object

\begin{equation} \label{eq:connection-def}
t_{\mu\nu}=\phi |\hat\Omega|^{1/2} q_{\mu\nu} \hspace{0.1cm}; \hspace{0.1cm} t^{\mu\nu}= \frac{q^{\mu\nu}}{\phi |\hat\Omega|^{1/2}} \ .
\end{equation}
Eq.(\ref{eq:connection3}) tells us that the independent connection, $\Gamma^{\lambda}_{\mu\nu}$, can be expressed as the Christoffel symbols of the rank-two tensor $t_{\mu\nu}$:

\begin{equation} \label{eq:LC}
\Gamma^\lambda_{\mu\nu}= \frac{t^{\lambda\alpha}}{2}\left(\partial_\mu t_{\alpha\nu}+\partial_\nu t_{\alpha\mu}-\partial_\alpha t_{\mu\nu}\right) \ .
\end{equation}
The reason to introduce this manipulation is that, as we shall see at once, the field equations in terms of $t_{\mu\nu}$ can be cast under a simple, Einstein-like representation.

Using again the definition $\hat\Omega={\hat g}^{-1}\hat q$ one can rewrite (\ref{eq:gvar}) as

\begin{equation}\label{eq:Omega-T}
\phi |\hat\Omega| {[{\hat\Omega}^{-1}]^\mu}_\nu=\frac{\left(\phi |\hat\Omega|+V(\phi)+\lambda\right)}{2}{\delta^{\mu}}_\nu-\frac{\kappa^2\epsilon}{2}{T^{\mu}}_\nu \ .
\end{equation}
Since we have already discussed that (\ref{eq:V_phi}) gives us $\phi=\phi(\vert \hat{\Omega} \vert)$, it follows that (\ref{eq:Omega-T}) simply establishes an algebraic relation between the matrix ${[{\hat\Omega}^{-1}]^\mu}_\nu$ and the energy-momentum tensor ${T^{\mu}}_\nu$. Once ${[{\hat\Omega}^{-1}]^\mu}_\nu$ is known, one can obtain an expression for $q_{\mu\nu}$ which only depends on $g_{\mu\nu}$ and the matter. This means that Eq.(\ref{eq:Omega-T}), rather than a differential equation involving second-order derivatives of the connection $\Gamma$, it is an equation depending linearly on $\Gamma$, on the first derivatives of $g_{\mu\nu}$ and on derivatives of $\vert \hat{\Omega} \vert$, which is just a function of the energy-momentum tensor of the matter. Therefore, the connection $\Gamma$ is a non-dynamical object completely determined by the matter.

Now, to cast the field equations in amenable form for calculations, we start from the basic definition $q_{\mu\nu}=g_{\mu\nu}+\epsilon R_{\mu\nu}(\Gamma)$ and multiply it by $q^{\mu\nu}$ to obtain $\epsilon q^{\mu\alpha} R_{\alpha\nu}(\Gamma)={\delta^{\mu}}_\nu-{ [\hat\Omega^{-1}]^{\mu}}_\nu$ . Next, using Eqs.(\ref{eq:connection-def}) and (\ref{eq:Omega-T}) we obtain

\be \label{eq:eom}
{R_\mu}^{\nu}(t)=\frac{\kappa^2}{2\phi^2 \vert \hat{\Omega} \vert^{3/2}} \left(L_G \delta_{\mu}^{\nu} + {T_\mu}^{\nu} \right) \ ,
\en
where ${R_\mu}^{\nu}(\Gamma) = {R_\mu}^{\nu}(t)$ since $\Gamma$ is the Levi-Civita connection of $t_{\mu\nu}$, see Eq.(\ref{eq:LC}), and the Lagrangian density reads explicitly

\be
L_G=\frac{1}{\kappa^2 \epsilon} \left(\phi \vert \hat{\Omega} \vert - V(\phi)-\lambda \right) \ .
\en
Note that, in general, $L_G$ and $\vert \hat{\Omega} \vert$ appearing on the right-hand-side of the field equations (\ref{eq:eom}) are functions of the matter sources. This means that (\ref{eq:eom}) represents a system of second-order field equations for $t_{\mu\nu}$, where all the objects on the right-hand-side depend solely on the matter sources. Solving these equations, and using the definitions (\ref{eq:connection-def}) and $q_{\mu\nu} = g_{\mu\alpha}{\Omega^\alpha}_{\nu}$ one can obtain $g_{\mu\nu}=t_{\mu\alpha}{[\hat\Omega^{-1}]^{\alpha}}_\nu/(\phi |\hat\Omega|^{1/2})$,  which provides a full solution to this problem, once a Lagrangian density, $f(\vert \hat{\Omega} \vert)$, and a matter source, $L_m$, are specified. Let us also point out that in vacuum, ${T_\mu}^{\nu}=0$, we can write $g_{\mu\nu}=t_{\mu\nu}$ (modulo a trivial re-scaling) and, together with the dependence of  ${\Omega_\alpha}^{\nu}$, $\vert \hat{\Omega} \vert$ and $\phi$ on the matter sources, the field equations (\ref{eq:eom}) boil down to those of GR (with a cosmological constant term as long as $\lambda \neq 1$), which implies the absence of ghost-like degrees of freedom.

\section{Electrovacuum spacetimes} \label{sec:III}

Let us consider now as the matter sector of our theory a standard electromagnetic (Maxwell) field, with Lagrangian density

\be
L_m=-\frac{1}{8\pi}F_{\mu\nu}F^{\mu\nu} \ ,
\en
where $F_{\mu\nu}=\partial_{\mu}A_{\nu}-\partial_{\nu}A_{\mu}$ is the field strength tensor of the vector potential $A_{\mu}$. Let us assume a static, spherically symmetric spacetime with line element, $ds^2=g_{tt}dt^2 - g_{rr}dr^2 -r^2 d\Omega^2$, where $r$ is the radial function. For an electrostatic, spherically symmetric field, whose only non-vanishing component is $F^{tr}(r)$, the Maxwell field equations, $\nabla_{\mu}(\sqrt{-g} F^{\mu\nu})=0$, lead to $F^{tr}=Q/(r^2 \sqrt{-g})$, where $Q$ is an integration constant identified as the electric charge. The corresponding energy-momentum tensor for this matter source reads

\bea \label{eq:Tmunu}
{T_\mu}^{\nu}= \frac{Q^2}{4\pi r^4}
\left(
\begin{array}{cc}
- \hat{I} &  \hat{0} \\
\hat{0} & \hat{I}  \\
\end{array}
\right) \ , \label{eq:em}
\ena
where $\hat{I}$ and $\hat{0}$ are the $2\times 2$ dimensional identity and zero matrices, respectively. In this case, the relation (\ref{eq:Omega-T}) reads

\be \label{eq:phi-omega-p}
\phi \vert \hat{\Omega} \vert {[{\hat\Omega}^{-1}]^\mu}_\nu = \frac{1}{2} \left(
\begin{array}{cc}
 \Omega_{-} \hat{I} &  \hat{0} \\
\hat{0} & \Omega_{+}  \hat{I}  \\
\end{array}
\right) \ ,
\en
where we have defined the objects

\be
\Omega_{\pm}= \phi \vert \hat{\Omega} \vert + V(\phi) + \lambda  \mp X \ ,
\en
and  $X \equiv \frac{\epsilon \kappa^2 Q^2}{4\pi r^4}$. Taking the determinant of (\ref{eq:phi-omega-p}) we obtain

\be \label{eq:trace}
\phi^2 \vert \hat{\Omega} \vert^{3/2}= \frac{\Omega_{-}\Omega_{+}}{4} \ ,
\en
which is an algebraic equation for the object $\vert \hat{\Omega} \vert$, which can be solved once the gravity Lagrangian density is given, at least numerically. On the other hand, the field equations read in this case

\be \label{eq:fe}
{R_\mu}^{\nu}(t)=\frac{2}{\epsilon \Omega_{-}\Omega_{+}} \left(
\begin{array}{cc}
\omega_{+} \hat{I} &  \hat{0} \\
\hat{0} & \omega_{-} \hat{I}  \\
\end{array}
\right) \ ,
\en
where we have defined $\omega_{\pm} = \phi \vert \hat{\Omega} \vert  -V(\phi)-\lambda \mp X$. These equations provide a full solution to this problem once the $f(A)$ function is specified. In the following section we shall consider a particular family of extensions of Born-Infeld gravity.

\section{A family of theories} \label{sec:IV}

Let us consider a family of theories of the form

\be
f(\vert \hat{\Omega} \vert)=\vert \hat{\Omega} \vert^{n/2} \ ,
\en
so that $n=1$ corresponds to the standard Born-Infeld gravity. For this family, the scalar field functions characterizing the model read

\be
A=\left(\frac{2\phi}{n}\right)^{\frac{2}{n-2}} \hspace{0.1cm} ; \hspace{0.1cm} V(\phi)=\frac{n-2}{2}\left(\frac{2\phi}{n}\right)^{\frac{n}{n-2}} \ .
\en
Using Eq.(\ref{eq:V_phi}) we obtain $\vert \hat{\Omega} \vert \equiv (2\phi/n)^{\frac{2}{n-2}}$ and therefore we get

\be  \label{eq:field-functions}
\phi=\frac{n}{2} \vert \hat{\Omega} \vert^{\frac{n-2}{2}} \hspace{0.1cm} ; \hspace{0.1cm}  V(\phi)=\frac{n-2}{2} \vert \hat{\Omega} \vert^{\frac{n}{2}} \ .
\en
The determinant equation (\ref{eq:trace}) becomes now

\be \label{eq:trace-p}
n^2 \vert \hat{\Omega} \vert^{\frac{2n-1}{2}}= \Omega_{-}\Omega_{+} \ ,
\en
where in this case we have

\be \label{eq:omega-p}
\Omega_{\pm}= (n-1)\vert \hat{\Omega} \vert^{\frac{n}{2}} + \sigma_{\pm} \ ,
\en
with the definition

\be
\sigma_{\pm}= \lambda \mp X \ .
\en
Eq.(\ref{eq:trace-p}) can be solved for any value of $n$, though numerical methods may be needed in some cases. This provides the fundamental object, $\vert \hat{\Omega} \vert$, that characterizes the theory and allows to solve the field equations (\ref{eq:fe}). Finally we need to write the explicit relation between $g_{\mu\nu}$ and $t_{\mu\nu}$ in this case, so we use (\ref{eq:phi-omega-p}) and (\ref{eq:field-functions}) to write the matrix:

\be
{\Omega_\mu}^{\nu}=n \vert \hat{\Omega} \vert^{\frac{n}{2}} \left(
\begin{array}{cc}
\Omega_-^{-1} \hat{I} &  \hat{0} \\
\hat{0} & \Omega_{+}^{-1}  \hat{I}  \\
\end{array}
\right) \ ,
\en
which implies

\bea \label{eq:t-g}
t_{\mu\nu}&=& \frac{n^2}{2} \vert \hat{\Omega} \vert^{\frac{2n-1}{2}} g_{\mu\nu} \left(
\begin{array}{cc}
\frac{1}{\Omega_{-}} \hat{I} &  \hat{0} \\
\hat{0} & \frac{1}{\Omega_{+}}   \hat{I}  \\
\end{array}
\right) \\
&=&\frac{1}{2} g_{\mu\nu} \left(
\begin{array}{cc}
\Omega_{+}\hat{I} &  \hat{0} \\
\hat{0} & \Omega_{-} \hat{I}  \\
\end{array}
\right) \ , \nonumber
\ena
where in the last equality Eq.(\ref{eq:trace-p}) has been used.

\subsection{Low-energy limit}

In the limit $\epsilon \ll 1$ (low-energy limit) one can formally write $\hat\Omega=\hat I+\epsilon \hat P$ and expand the determinant in series of $\epsilon$  as $|\hat\Omega|\approx 1+\epsilon \text{Tr}[\hat P]+\frac{\epsilon^2}{2}\left( \text{Tr}[\hat P]^2-\text{Tr}[\hat P^2]\right)+O(\epsilon^3)$, where $\text{Tr}[T]$ denotes the trace of $T$. Since one can verify that $\text{Tr}[\hat P]=R$ and $\text{Tr}[\hat P^2]=R_{\mu\nu}R^{\mu\nu}$, it follows that

\begin{equation}
\lim_{\epsilon\to 0} |\hat\Omega|^{n/2}\approx 1+\frac{n\epsilon}{2} R+\frac{n\epsilon^2}{4}\left( \frac{n}{2} R^2-R_{\mu\nu}R^{\mu\nu}\right)+O(\epsilon^3) \ .
\end{equation}
Therefore, the action in the low-energy limit boils down to

\begin{eqnarray} \label{eq:action-low}
\lim_{\epsilon\to 0}S_f&=&\int d^4 x\sqrt{-g}\left[\frac{(1-\lambda)}{\epsilon\kappa^2}+\frac{n}{2\kappa^2}R \right] \nonumber \\
&+& \epsilon \int d^4x \sqrt{-g} \left[ \frac{n }{4\kappa^2}  \left( \frac{n}{2} R^2-R_{\mu\nu}R^{\mu\nu}\right) \right] \\
&+& O(\epsilon^2) +  S_m \ . \nonumber
\end{eqnarray}
With the definitions $\tilde{\kappa}^2\equiv \kappa^2/(n)$ and $\Lambda=(\lambda-1)/(n\epsilon)$ we can read the action (\ref{eq:action-low}) as that of Einstein-Hilbert with a cosmological constant term $\Lambda$, and the next-to-lowest order defines a quadratic gravity model. Since the parameter $\epsilon$ has dimensions of length squared, one could assume, from quantum gravity considerations, to be of order $\epsilon \sim l_P^2$, where $l_P = \sqrt{\hbar G/c^3}$ is Planck's length. Since the limit $\epsilon \rightarrow 0$ represents situation where the curvatures involved (or the energy scales) are much smaller than Planck's scale, this implies that the predictions of these theories are in completely agreement with all available observational data (in particular, those coming for solar system constraints, see e.g. \cite{LCDM}). However, on the high-energy region, where the curvatures involved are extremely large (like in the innermost region of black holes), they provide theoretical scenarios where we can probe new dynamics beyond GR, which is precisely the regime we are interested in this work.

\subsection{Solution of the field equations}

To solve the field equations (\ref{eq:fe}) with the definitions above we introduce two line elements imposing staticity and spherical symmetry, one for $g_{\mu\nu}$:

\be \label{eq:ds2}
ds^2=-A(r)dt^2  +B(r)^{-1} dr^2  +r^2 d\Omega^2 \ ,
\en
and another for $t_{\mu\nu}$:

\be
d\tilde{s}^2=-e^{2\psi(x)}C(x) dt^2 + C(x)^{-1}dx^2 + x^2 d\Omega^2 \ ,
\en
where the two sets of functions $(A(r), B(r))$ and $(e^{2\psi}(x), C(x))$ are to be determined by resolution of the field equations and the transformations (\ref{eq:t-g}), and $d\Omega^2$ is the angular sector on each line element. In particular, the transformations (\ref{eq:t-g}) yield the relation between the radial coordinates of each metric as

\be \label{eq:r-rtilde}
x^2=\frac{1}{2}r^2 \Omega_{-} \ ,
\en
which implies that $r=r(x)$ in Eq.(\ref{eq:ds2}). This relation plays a key role in the characterization of the innermost structure of the corresponding solutions, as we shall see at once.

Now, using Mathematica and the fact that the components of the independent connection $\Gamma^{\lambda}_{\mu\nu}$ are given by the Christoffel symbols of $t_{\mu\nu}$, see Eq.(\ref{eq:LC}), we obtain

\begin{eqnarray}\label{eq:Rtt}
{R_t}^t&=&-\frac{C_{xx}}{2}\Bigg[\frac{C_{xx}}{C}-\left(\frac{C_{x}}{C}\right)^2+
2\psi_{xx}+ \nonumber \\ &+& \left(\frac{C_{x}}{C}+ 2\psi_{x}\right)\left(\frac{C_{x}}{C}+
\psi_{x}+\frac{2}{x}\right)\Bigg] \\
{R_{x}}^{x}&=&-\frac{C_{xx}}{2}\Bigg[\frac{C_{xx}}{C}-
\left(\frac{C_{x}}{C}\right)^2+2\psi_{xx}+ \nonumber \\ &+& \left(\frac{C_{x}}{C}+
2\psi_{x}\right)\left(\frac{C_{x}}{C}+\psi_{x}\right)+
\frac{2}{x}\frac{C_{x}}{C}\Bigg] \label{eq:Rrr}\\
{R_\theta}^\theta&=&\frac{1}{x^2}\left[1-C(1+x\psi_{x})-xC_{x}\right] \ . \label{eq:Rzz}
\end{eqnarray}
where $\psi_x \equiv \frac{d\psi}{dx}$, $C_x \equiv \frac{dC}{dx}$ and so on. From the substraction ${R_t}^t-{R_{x}}^{x}=\psi_{x}$, and equating to the right-hand-side of the field equations (\ref{eq:fe}), it follows that $\psi_{x}=0$ and thus $\psi=$constant, that can be put to zero by a redefinition of the time coordinate, like in GR. To solve the component $({_\theta}^{\theta})$ of the field equations, we introduce a mass ansatz

\be
C(x)=1-\frac{2M(x)}{x} \ ,
\en
which, inserted into (\ref{eq:fe}) together with the definitions of this section, yields

\be \label{eq:Mrtilde}
M_{x}=x^2\frac{\vert \hat{\Omega} \vert ^{n/2} - \sigma_{+}}{\epsilon n^2 \vert \hat{\Omega} \vert^{\frac{2n-1}{2}}} \ ,
\en
where $M_x \equiv \frac{dM}{dx}$ and we have used the relation (\ref{eq:r-rtilde}). Taking a derivative in (\ref{eq:r-rtilde}) we obtain

\be \label{eq:dxdr}
\frac{dx}{dr}= \pm \frac{(n-1)\vert \hat{\Omega} \vert^{\frac{n}{2}} (1+\frac{n}{4} \frac{r\vert \hat{\Omega} \vert_r}{\vert \hat{\Omega} \vert}) + \sigma_{+}}{\sqrt{2}\Omega_{-}^{1/2}} \ ,
\en
Using this in (\ref{eq:Mrtilde}) we get the result

\bea
M_r&=&\frac{r^2}{2}\frac{(\vert \hat{\Omega} \vert^{\frac{n}{2}}-\sigma_{+})\Omega_{-}^{1/2}}{\epsilon n^2 \vert \hat{\Omega} \vert^{\frac{2n-1}{2}}}\\
&\times& \Big[ (n-1)\vert \hat{\Omega} \vert^{\frac{n}{2}} \Big(1+\frac{n}{4} \frac{r\vert \hat{\Omega} \vert_r}{\vert \hat{\Omega} \vert} \Big)+\sigma_{+} \Big] \ ,
\ena
where $M_r \equiv \frac{dM}{dr}$. Therefore, the function $C(r)$ gets determined by

\be \label{eq:A-final}
C(r)=1-\frac{2M(r)}{r \Omega_{-}^{1/2}} \ ,
\en
where we stress that $r=r(x)$. To simplify the notation, let us introduce some definitions. First, since in this work we are interested in the case $\epsilon<0$ (to match with the results of \cite{ors15b,ors15}, where wormhole and regular solutions were found) we define the length scale $l_{\epsilon}^2=-\vert \epsilon \vert /2$ (note that this induces a change in the sign of $X$ and, consequently, in the related objects $\sigma_{\pm}$ and $\Omega_{\pm}$). In addition, we introduce a new dimensionless radial coordinate $z=r/r_c$, where $r_c^4= l_{\epsilon}^2 r_q^2$, with $r_q^2 = \kappa^2 Q^2/(4\pi)$ (so $X=-1/z^4$). In these units, we can formally integrate the mass function as $M(z)=M_0 (1+\delta_1 G(z))$ (where $r_S=2M_0$ is Schwarzschild's radius, which emerges as an integration constant) and write

\be \label{eq:A-parameterized}
C(z)=1-\frac{1+\delta_1 G(z)}{\delta_2 z \Omega_{-}^{1/2}} \ ,
\en
so now $z=z(x)$. This way the problem gets parameterized by the two constants

\be
\delta_1=\frac{1}{2r_S} \sqrt{\frac{r_q^3}{l_{\epsilon}}} \hspace{0.1cm} ; \hspace{0.1cm} \delta_2=\frac{\sqrt{r_q l_{\epsilon}}}{r_S} \ ,
\en
while the (dimensionless) function $G(z)$ satisfies

\bea
\frac{dG}{dz}&=&-z^2\frac{(\vert \hat{\Omega} \vert^{\frac{n}{2}}-\sigma_{+})\Omega_{-}^{1/2}}{n^2 \vert \hat{\Omega} \vert^{\frac{2n-1}{2}}} \nonumber \\
&\times& \Big[ (n-1)\vert \hat{\Omega} \vert^{\frac{n}{2}} \Big(1+\frac{n}{4} \frac{z \vert \hat{\Omega} \vert_z}{\vert \hat{\Omega} \vert} \Big)+\sigma_{+} \Big] \ . \label{eq:diffg}
\ena
Finally, we make use of the transformation (\ref{eq:t-g}) and take into account Eq.(\ref{eq:r-rtilde}) to write the line element (\ref{eq:ds2}) for the physical metric $g_{\mu\nu}$ as

\be\label{eq:line-final}
ds^2=-Adt^2 + \frac{1}{A\Omega_{+}^2} dz^2 + z^2(x) d\Omega^2 \ ,
\en
where $A=C/\Omega_{+}$. When $n=1$ (standard Born-Infeld gravity case) the metric expression gets simplified as the function $\vert \hat{\Omega} \vert$ can be analytically solved as $\vert \hat{\Omega} \vert=\sigma_{+}^2 \sigma_{-}^2$ and it follows that the function $G(z)$ satisfies now
\be
G_z=\left( \frac{1}{z^2} - \frac{\lambda-1}{\lambda} z^2 \right) \frac{\sigma_{+}}{\sigma_{-}^{1/2}} \ ,
\en
which reduces to expression $G_z=\frac{z^4+1}{z^4(z^4-1)^{1/2}}$ in the asymptotically flat case. The corresponding properties of these solutions were analyzed in detail in Ref.\cite{ors}. In vacuum, $M(x)=M_0$, and the solutions reduce to the Schwarzschild black hole of GR. In what follows we shall only consider asymptotically flat spacetimes, $\lambda=1$.

Due to the Maxwell fall off of the electromagnetic field, at large distances we have $z^2(x) \simeq x^2$, $G(z) \simeq -1/z$ and these geometries recover the Reissner-Nordstr\"om (RN) solution of the Einstein-Maxwell field equations, $A(r) \simeq 1-r_S/r+r_q^2/(2r^2)$. This implies, in particular, that the locations of the horizons, given by the solutions $z_h$ of the function $A(z_h)=0$, which is translated into the equation $\delta_2 z_h \Omega_{-}(z_h)^{1/2}=1+\delta_1 G(z_h)$, will much be the same as those of the RN solution, at least for macroscopic black holes. Only as the region of high-density or very short scales (of size $\sim l_{\epsilon}^2$) is approached, one finds deviations with respect to those solutions, and this is actually the region we are interested in this work. The most remarkable modification comes from the behaviour of the radial coordinate $z(x)$. For $z \gg 1$ this relation is linear so the role of $x$ as the standard radial coordinate of GR is restored [as so is the RN behaviour]. In this sense, we note that using the coordinate $z(x)$ in the line element (\ref{eq:line-final}) is under the important constraint of being strictly monotonic, and thus the change of coordinates (\ref{eq:dxdr}) could be not well defined beyond $x=0$ (where $z=1$) in those cases where $\vert \hat{\Omega} \vert_{z=1}=0$, since it can induce the behaviour $dz/dx=0$ there. In such a case, to cover the whole spacetime one would need a single coordinate $x \in (-\infty,\infty)$, but two copies of the coordinate $z$, one for the interval where $z$ grows with growing $x$, and another for the interval where $z$ decreases with growing $x$, which is reflected on the existence of two signs in Eq.(\ref{eq:dxdr}), namely

\begin{equation} \label{eq:coc1}
\frac{x}{r_c}= \pm z \Omega_{-}^{1/2} \ ,
\end{equation}
where we have reabsorbed the factor $\sqrt{2}$ by redefining $x \rightarrow \sqrt{2} x$. It is thus clear that the radial coordinate $z(x)$ reaches a minimum at $x=0$, where it bounces off. There is an immediate and well known interpretation for this fact, which is that the line element (\ref{eq:line-final}) describes a wormhole structure [see \cite{Visser} for a full account on wormhole physics], where the throat $z=1$ ($x=0$) interpolates between two asymptotically flat space-time regions. Nonetheless, the possibility of such an extension beyond $x=0$ is subjected to some constraints, as we shall see below. Wormholes are non-trivial topological structures, which in GR are unavoidable supported by \emph{exotic} energy-matter sources violating the energy conditions \cite{Visser}, though this does not need to be so in the context of modified gravity.

To further describe the wormhole geometry we would need the explicit expression of $z=z(x)$, which requires to invert the relation (\ref{eq:coc1}) and thus the full knowledge of the object $\vert \hat{\Omega} \vert$ is needed. This is a highly non-trivial task, though, and the obtention of a closed expression for this object in all the range of definition of the solutions seems not to be possible (at least, analytically). Therefore, in the next section we will consider series expansions around $z=1$, which nonetheless is the region we are interested, and study the behaviour of the metric functions there.

\subsection{Finding $|\hat \Omega|$}

To simplify the analysis, let us focus on asymptotically flat solutions, $\lambda=1$. Inserting Eq.(\ref{eq:omega-p}) into Eq.(\ref{eq:trace-p}), taking the square root on both sides and with a little rearrangement we get
\be \label{eq:trace-expansion}
n^2 |\hat\Omega|^{n-\frac{1}{2}}-(n-1)^2 |\hat\Omega|^{n}-2(n-1)|\hat\Omega|^{\frac{n}{2}}=1-z^{-8} \ .
\en
This is the key equation to be solved for a given $n$ around the region $z=1$, since this is the minimum value accessible to the radial function. To this end we write $|\hat \Omega|$ in terms of power series over that region as
\begin{equation} \label{eq:Om}
|\Omega|=\sum_{i=1}^{\infty}a_{i}(z-1)^{n_{i}} \ ,
\end{equation}
where $a_i$ and $n_i$ (which are not necessarily integers) are constants depending only on $n$, which labels each theory. By inserting the expression (\ref{eq:Om}) into the left-hand-side of Eq.(\ref{eq:trace-expansion}), and performing series expansion on the right-hand-side, we just need the coefficients of each term on both sides to be equal. A glance at the different leading terms and their powers on both sides of the Eq.(\ref{eq:trace-expansion}) naturally leads to a classification in three cases: $0<n < 1/2$, $1/2< n \leq 1$ and $n>1$, which we study separately.

\subsubsection{$0<n < \frac{1}{2}$}

In this case the lowest order term on the left-hand side is $n^2 |\hat\Omega|^{n-\frac 12}$. However, as the region $z=1$ is approached, the left-hand side of Eq.(\ref{eq:trace-expansion}) would be divergent when $|\hat \Omega| \rightarrow 0$, while the right-hand-side vanishes. Thus this case cannot represent wormhole solutions and we disregard it from our analysis.

\subsubsection{$\frac{1}{2} < n\leq 1$}

\begin{figure}[h]
\includegraphics[width=8.5cm,height=5cm]{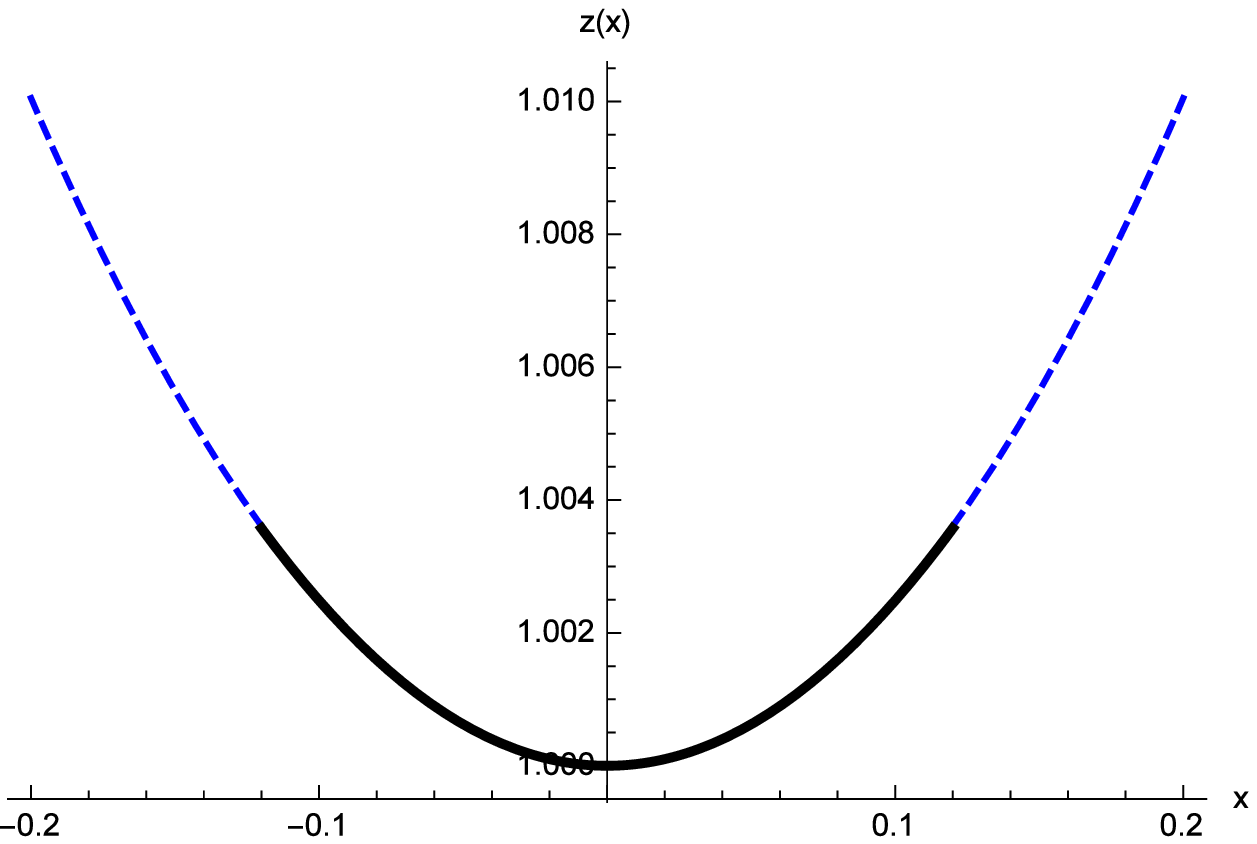}
\includegraphics[width=8.5cm,height=5cm]{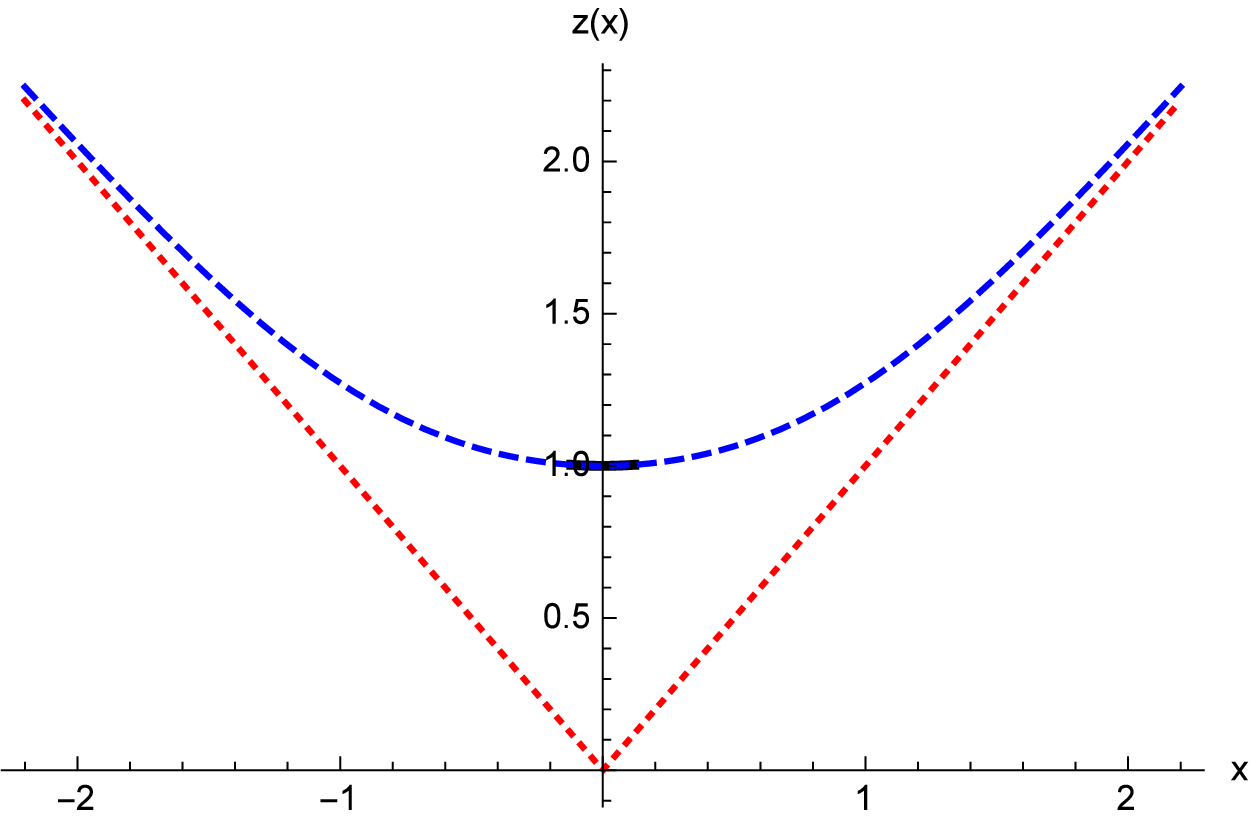}
\caption{Top figure: representation of the radial function $z(x)$ around the wormhole throat $x=0$ in the case $1/2<n \leq 1$ (solid black line), resulting from solving Eq.(\ref{eq:coc1}) with the approximation (\ref{eq:WHappn}). The dashed blue line represents the behaviour found in the standard Born-Infeld gravity model ($n=1$) in the full range, and both solutions coincide in the region around $x=0$, as expected. Bottom figure: we have enhanced the range of $x$ to compare with the GR solution (dotted red line), $z = \vert x \vert $, to which all solutions with $1/2<n \leq 1$ approach (because of the Maxwell fall off in all cases) for large $x$.\label{fig:1}}
\end{figure}

In this case the lowest order term on the left-hand side is still $n^2 |\hat\Omega|^{n-1/2}$, which goes to zero as $|\hat\Omega| \rightarrow 0$, which is consistent with the behaviour of the right-hand-side. Solving the corresponding equation for the leading order term one gets
\be\label{eq:Om1/2}
|\hat\Omega| \simeq\left(\frac {8}{n^2}\left(z-1\right)\right)^{\frac 2{2n-1}}+ \ldots \ ,
\en
where hereafter dots refer to other terms than the dominant ones. Eq.(\ref{eq:On1}). Replacing this expression in the definition of $\Omega_{-}$, given in Eq.(\ref{eq:omega-p}), we obtain
\be \label{eq:Omn}
\Omega_{-} \simeq 4(z-1)+\left(\frac{8(z-1)}{n^2}\right)^{\frac {n}{2n-1}}+ \ldots \ .
\en
Note that, since $\frac{1}{2}<n \leq 1$ the $(z-1)$ term is actually the leading order term in Eq.(\ref{eq:Omn}). As $\Omega_{-}$ is the key object governing the behaviour close to the region $z=1$ [see Eq.(\ref{eq:coc1})], this implies a universal behaviour for all the models within this subclass, which is exactly the same as that found in the original Born-Infeld gravity model ($n=1$). We note that using Eq.(\ref{eq:coc1}) with the approximation (\ref{eq:Omn}) yields a cubic equation

\be \label{eq:WHappn}
x^2 \simeq 4z^2(z-1) \ ,
\en
for the behaviour of the radial function $z=z(x)$ around the wormhole throat. Solving this equation numerically we plot $z(x)$ in Fig.\ref{fig:1} around the wormhole throat $z=1$, and compare it with the known behaviour for standard Born-Infeld gravity ($n=1$). As expected, we find a complete agreement around the throat $x=0$, this result due to the same behaviour there of the fundamental object characterizing the theory and its solutions, $|\hat\Omega|$. Thus, a smooth bounce occurs at $x=0$, which confirms the presence of a wormhole structure in all these cases \footnote{For completeness of this analysis, we point out that in the case $n=1/2$, performing similar expansions one gets that around $x=0$ the relation between coordinates is expressed as $x^2=\frac{z^3}{2} +O(z^5)$. This implies that $\frac{dz}{dx} \vert_{x=0}$ and, as we shall below, these cases cannot be identified as wormholes.}.

\begin{figure}[h]
\includegraphics[width=8.5cm,height=6cm]{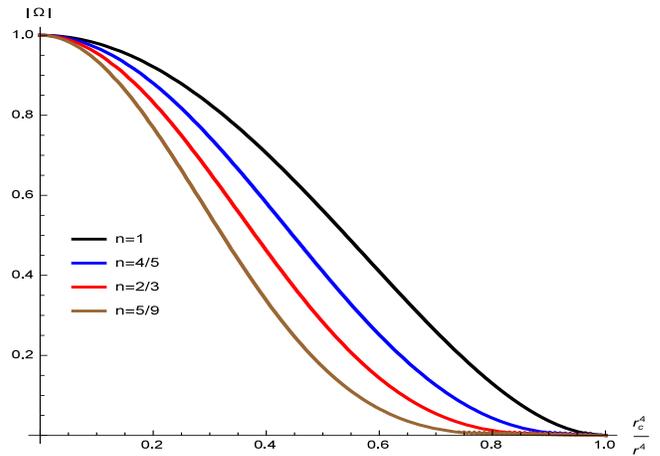}
\caption{Representation of $|\hat\Omega|$ as a function of $r_c^4/r^4$ when $1/2<n \leq 1$ for the values $n=1$ (solid black curve) corresponding to standard Born-Infeld gravity, and the models $n=4/5, 2/3, 5/9$. \label{fig:2b}}
\end{figure}

We also note that the fundamental object $|\hat\Omega|$ can be numerically computed in all range of definition of $z \in [1,\infty)$, and is plotted in Fig.\ref{fig:2b}. There it is seen that for any value of $\frac{1}{2}<n \leq 1$ the object $|\hat\Omega|$ interpolates smoothly and uniquely between asymptotic infinity, $r_c^4/r^4 \rightarrow 0$, where $|\hat\Omega|=1$, and the wormhole throat, $r^{4}_c/r^4=1$, where $|\hat\Omega|=0$.

\subsubsection{$n>1$} \label{sec:IIIC3}

In this case the leading term on the left-hand side of (\ref{eq:trace-expansion}) is $-2(n-1)|\hat\Omega|^{\frac{n}{2}}$, which has the correct vanishing behaviour as $ |\hat\Omega| \rightarrow 0$. Equating both sides of Eq.(\ref{eq:trace-p}) we obtain

\be
|\hat\Omega|=\left(\frac{4(1-z)}{n-1}\right)^{\frac 2n}+\mathcal{O}(z-1)^{\frac 2n} \ .
\en
To get the expression for $\Omega_{-}$ in Eq.(\ref{eq:omega-p}), we find that when replacing the leading order term it cancels out with the $\sigma_{-}$ contribution and, therefore, we have to calculate the next-to-leading order of $|\hat\Omega|$ to get a consistent result. This is done by requiring the next-to-leading order term in $-2(n-1)|\hat\Omega|^{\frac{n}{2}}$ to cancel with the leading term of $n^2 |\hat\Omega|^{n-1/2}$ in Eq.(\ref{eq:trace-expansion}). So finally we have

\be \label{eq:On1}
\Omega_{-} \simeq \frac {n^2}{2}\left(\frac {4(1-z)}{n-1} \right)^{2- \frac{1}{n}} +\ldots \ ,
\en
which is the result we need for the computations of the next section.

\section{Characterization of the solutions} \label{sec:V}

Once the explicit behaviour of $|\hat\Omega|$ and $\Omega_{-}$  around the relevant region $z=1$ ($x=0$) are known, we can study the metric there by inserting such expressions into Eqs.(\ref{eq:diffg}) and (\ref{eq:A-parameterized}). Consistently with the previous section, let us split our analysis into two separated cases

\subsection{$1/2<n \leq 1$}

In this case, to lowest order in $(z-1)$, we have the following behaviour for the function $G(z)$ in (\ref{eq:diffg})

\bea
\frac{dG}{dz}& \simeq &\frac{1}{(z-1)^{\frac{1}{2}}}+ \ldots \\
G(z)& \simeq &-1/\delta_c^{n}+2(z-1)^{\frac{1}{2}} +\ldots \ .
\ena
where $\delta_c^{n}$ is an integration constant for each model labelled by $n$, whose value must be properly tuned to give the right far contribution (Coulombian) of the electromagnetic field at large distances, $G(z) \simeq -1/z$, of the RN solution of GR. With the expressions above, the corresponding expansion for the metric function $A(x)$ becomes (note that $\Omega_{+} \simeq 2$ around $z=1$).

\be
A(z) \simeq \frac{\delta_1-\delta_c^{(n)}}{4\delta_2 \sqrt{z-1}} + \frac{1}{2} \left(1-\frac{\delta_1}{\delta_2} \right) + \ldots  \ . \label{eq:A-1/2}
\en
This way, we have all the necessarily elements to characterize the behaviour of the geometry (\ref{eq:line-final}) around $z=1$ (or $x=0$). We already know, from the bouncing of the radial function $z(x)$ around the region $x=0$ displayed in Fig.\ref{fig:1}, that all solutions with $1/2<n <1$ represent a wormhole structure with exactly the same behaviour as the solutions of the standard Born-Infeld gravity extension of GR (corresponding to $n=1$). This is in sharp contrast to the GR behaviour (Reissner-Norsdstr\"om solutions) where, instead of a wormhole, as one approaches to $r=0$ a point-like singularity is met. To further understand the geometry there, we can compute the behaviour of the Kretschman scalar, $K={R_\alpha}^{\beta\gamma\delta}{R^\alpha}_{\beta\gamma\delta}$, at the wormhole throat $x=0$, which yields a general structure

\bea
K(z) &\simeq & \frac{(\delta_1-\delta_c^{(n)})^3}{4\delta_2^2 (\delta_c^{(n)})^2(z-1)^3} - \frac{5(\delta_1-\delta_c^{(n)})^2}{8\delta_2^2 (\delta_c^{(n)})^2(z-1)^2} \nonumber \\
 &+& \mathcal{O}\left(\frac{(\delta_1-\delta_c^{(n)})}{(z-1)} \right) \ .
\ena
Thus, the Kretchsman generically diverges there, though such a divergence can be removed for the particular choice $\delta_1=\delta_c^{(n)}$, since in this case one gets the result

\be \label{eq:Kr}
K(z) \simeq \gamma_n + \eta_n (z-1) + \ldots \ ,
\en
where the constants $\gamma_n$ and $\eta_n$ take complicated expressions depending both on the value of $n$ chosen and how many additional terms in the series (\ref{eq:A-1/2}) are taken, though the general structure (\ref{eq:Kr}) remains unmodified. Nonetheless, we point out that the existence of divergences in (some of) the curvature scalars does not necessarily imply the existence of a spacetime singularity, despite the fact that in much of the literature both concepts are taken to be equivalent (see e.g. \cite{Ansoldi}, where different solutions free of curvature divergences are reviewed and summarized). This is so because in the formulation of the singularity theorems \cite{theorems} no reference is given to pathologies in the curvature scalars, but instead they make use of the concept of geodesic completeness, namely, whether any time-like or null geodesic can be extended to arbitrarily large values of the affine parameter or not. Indeed, this is widely accepted to be the most reliable criterium to deal with spacetime singularities \cite{Books}, since time-like and null geodesics are associated to the free falling paths of physical observers and to the transmission of information, respectively [see \cite{Senovilla} for a recent discussion on this topic]. On the other hand, the effect of curvature divergences upon extended observers crossing the divergent region is a question that has to be analyzed separately \footnote{This has investigated in detail in the case of standard Born-Infeld gravity ($n=1$), with the result that the presence of curvature divergences do not necessarily imply the loss of causal contact among the constituents making up a physical observer \cite{congruences}, and that the problem of scattering of waves off the wormhole is well posed \cite{ors15b}, supporting the conclusion that no absolutely destructive effect would happen upon a physical observer crossing the wormhole throat.}.

\begin{figure}[h]
\includegraphics[width=7.5cm,height=5cm]{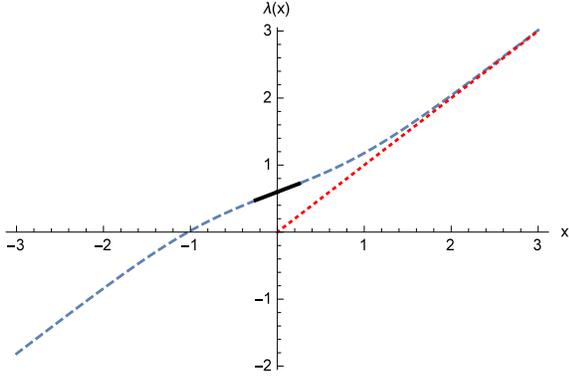}
\caption{Affine parameter $\lambda$ as a function of the radial coordinate $x$ for radial null geodesics and $E=1$. The tiny solid black line around $x=0$ corresponds to the class of models with $1/2<n \leq1$, which holds only around the wormhole throat [see the approximations employed in the text], while the dashed blue line corresponds to the standard Born-Infeld gravity ($n=1$) in full range. The (straight) dotted red line corresponds to the GR geodesics, $\lambda(x)=x$, which are only defined for $x>0$. It is immediately seen that the affine parameter $\lambda(x)$ can be smoothly extended across the wormhole throat for any model $1/2<n \leq1$. \label{fig:2}}
\end{figure}

To investigate the geodesic completeness issue in our case we consider a geodesic curve, $\gamma^{\mu}=x^{\mu}(\lambda)$, with tangent vector $u^{\mu}=dx^{\mu}/d\lambda$, where $\lambda$ is the affine parameter and, instead of solving the geodesic equation, $d^2x^{\mu}/d\lambda^2+ \Gamma_{\alpha\beta}^{\mu}(dx^{\alpha}/d\lambda) (dx^{\beta}/d\lambda)=0$, we consider the norm of the tangent vector, $u_{\mu}u^{\mu}=-k$, where $k=0(1)$ for null(time-like) geodesics. This is advantageous due to the large amount of symmetry present in our problem, which allows us to rotate the plane to make it coincide with $\theta=\pi/2$ without loss of generality, and to identify two conserved quantities, namely, $E=A dt/d\lambda$ and $L=r^2 d\varphi/d\lambda$ [see \cite{ors15b} for a more detailed explanation of this point]. Thus, with the line element (\ref{eq:line-final}) the geodesic equation reads

\be \label{eq:geo}
-A\left(\frac{dt}{d\lambda} \right)^2 + \frac{1}{A(x)\Omega_{+}^2} \left(\frac{dx}{d\lambda} \right)^2 + r^2(x) \left(\frac{d\varphi}{d\lambda} \right)^2=-k \ ,
\en
and can be rewritten in terms of the conserved quantities as

\be \label{eq:geoc}
\frac{1}{\Omega_{+}^2} \left(\frac{dx}{d\lambda} \right)^2=E^2-A \left(k+\frac{L^2}{r(x)^2} \right) \ .
\en
In the Reissner-Nordstr\"om solution of the Einstein-Maxwell field equations one finds that null radial geodesics ($k=L=0$) are \emph{incomplete}, namely, they cannot be extended to arbitrarily large values of the affine parameter, but instead finish at $r=0$ [where they meet the curvature divergence] in a finite affine time with no possibility of further extension beyond this point [because $r>0$] hence, the singular character of that spacetime [see e.g. Ref.\cite{Chan} for a full description of this case]. In our case it is convenient to rewrite the equation (\ref{eq:geo}) in terms of the radial coordinate $r$ using Eq.(\ref{eq:dxdr}), which for null radial geodesics yields the result

\be
\frac{1}{\Omega_{+}^2} \frac{\left[(n-1)\vert \hat{\Omega} \vert^{\frac{n}{2}} \left(1+\frac{n}{4} \frac{z \vert \hat{\Omega} \vert_z}{\vert \hat{\Omega} \vert} \right) + \sigma_{+} \right]^2}{\Omega_{-}} \left(\frac{dr}{d\lambda} \right)^2 =E^2 \ .
\en
Around the throat, $x=0$, using the approximations (\ref{eq:Om1/2}) and $\Omega_{+} \simeq 2$ one gets the result $\frac{1}{\Omega_{-}} \left(\frac{dr}{d\lambda} \right)^2\simeq E^2$, whose integration can be conveniently written as

\be
\pm E \lambda(x) \simeq \int \frac{dz}{\Omega_{-}(z)^{1/2}} = \left\{\begin{array}{lr} x_0 + \sqrt{z-1} & \text{ if } x>0 \\  & \\
x_0 - \sqrt{z-1} & \text{ if } x<0 \end{array}\right\} \ .
\en
where $x_0 \simeq 0.59907$ is a constant. In Fig.\ref{fig:2} we have plotted the behaviour of these geodesics for $E=1$ around the wormhole throat, where the approximations above are valid. In the GR case (dotted red line) one has $\lambda(x)=x$ and, since in that case $x>0$, one finds a geodesically incomplete spacetime. However, the deviations introduced by the new dynamics of the models with $1/2<n \leq 1$ and, in particular, the  emergence of the wormhole, allow null radial geodesics to be complete. This is in agreement with the fact that the function $\Omega_{-}$ encoding both the wormhole structure and the geodesic equation takes exactly the same form as that of the standard Born-Infeld gravity ($n=1$), which has been shown in \cite{ors15} to be geodesically complete for these null radial geodesics. Moreover, since standard Born-Infeld gravity has also been shown to be geodesically complete for time-like and null geodesics with $L \neq 0$, and given the fact that the presence of angular momentum in (\ref{eq:geoc}) introduces an additional $r^2(x)$-dependent term [but which we recall has the same behaviour around the throat in all the $1/2<n \leq 1$ models], the same complete behaviour for these geodesics will hold for the models considered in this section.

\subsection{$n>1$}

Using the corresponding equations and discussion of section (\ref{sec:IIIC3}), to lowest order in $(z-1)$ now one gets the result
\bea
\frac{dG}{dz}&=\frac{4\sqrt2}{n}(\frac{4}{n-1})^{\frac {1}{2n}-1}(1-z)^{\frac {1}{2n}-1}+\mathcal{O}(z-1)^{-1}\\
G(z)&=-1/\delta_c^{n}+8\sqrt2 (\frac{4}{n-1})^{\frac {1}{2n}-1}(1-z)^{\frac {1}{2n}}+\mathcal{O}(1) \ ,
\ena
where $\delta_c^{(n)}=-1/\beta_2(n)$, and $\beta_2(n)$ arises as an integration constant for each model, similarly defined as in the cases $1/2<n \leq 1$. Inserting this into the definition of $A(z)$ yields

\bea
A(z) &\simeq & \frac{(\delta_1-\delta_c^{(n)})}{2\delta_2 \delta_c^{(n)}\xi(n)(z-1)^{\frac{2n-1}{2}}} \\ \nonumber
&+& \frac{1}{2} -\frac{\delta_1 \eta(n)}{\delta_2} (1-z)^{\frac{1-n}{n}} + \ldots \ ,  \label{eq:A-1}
\ena
where $\xi(n)=\frac{n}{\sqrt{2}} (\frac{4}{n-1})^{\frac{2n-1}{2n}}$ and $\eta(n)=\frac{8}{n} (4/(n-1))^{\frac{1-2n}{n}}$ are two constants. In this case, computation of the Kretschman scalar yields the generic divergent result

\be \label{eq:Knl1}
K \simeq \frac{a(\delta_1-\delta_c^{(n)})^2}{(z-1)^{2(n+1)}} + \ldots \ ,
\en
($a$ a constant) which cannot be completely removed even for the choice $\delta_1=\delta_c^{(n)}$ (unlike in the previous case), due to the existence of many additional divergent terms in the expression (\ref{eq:Knl1}) whose factor structure is not of the form $(\delta_1-\delta_c^{(n)})$.

\begin{figure}[h]
\includegraphics[width=7.5cm,height=5cm]{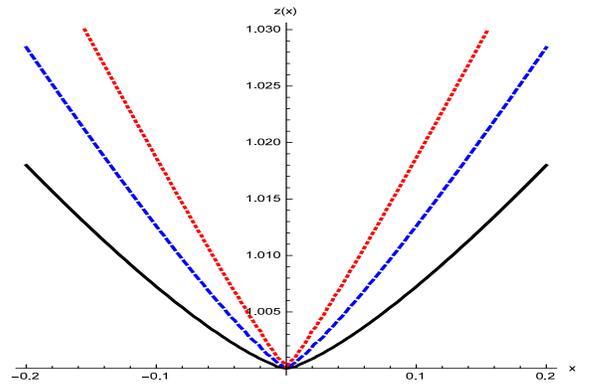}
\caption{The radial function $z(x)$ around the wormhole throat for the models with $n=2$ (solid black), $n=3$ (dashed blue) and $n=5$ (dotted red). In this plot we have used the approximation (\ref{eq:On1}). As opposed to the $1/2<n<1$ cases, the bounce in the radial coordinate is not completely smooth since it fails to satisfy the condition $\frac{dz}{dx}\vert_{x=0} = 0$ and thus these cases are unlikely to be interpreted as genuine wormholes. \label{fig:4}}
\end{figure}

In any case, the important point here is to realize that, though the radial coordinate bounces off at $x=0$ [see Fig.\ref{fig:4}], now it satisfies that $\frac{dz}{dx}\vert_{x=0} \neq 0$, as opposed to the cases $1/2<n \leq 1$, where one has $\frac{dz}{dx}\vert_{x=0}=0$. The relevance of this statement lies on the fact that, in order to have a well defined wormhole geometry, the latter condition has to be satisfied (see \cite{Visser}, chapter 11.2). Thus, the lack of regularity of the bounce and of a well defined wormhole geometry makes a geodesic completeness analysis to be ill defined. As we were interested in looking for regular solutions, we shall not proceed further with the analysis of the corresponding geometries.

\section{Conclusions} \label{sec:VI}

In this work we have worked out a formalism for studying black hole spacetimes in the context of functional extensions of Born-Infeld gravity, motivated by the fact that such a theory has been recently shown to be able to remove spacetime singularities. These extensions are formulated in the Palatini approach, where metric and connection are regarded as independent degrees of freedom. We have shown that they satisfy a set of second-order field equations that recover the GR equations in vacuum (with a cosmological constant term, in general) and are consequently free of ghosts. This analysis complements other extensions of Born-Infeld gravity recently considered in the literature \cite{Bou}, whose interest has greatly increased due to the many applications in astrophysics and cosmology that these theories have.

In this work we have focused on the innermost structure of electrovacuum black hole solutions, and studied in detail a family of power-law models labelled by a single parameter, $n$. Such a family is introduced mainly due to two reasons: first, recovery of GR plus quadratic corrections in the low-energy limit, $\epsilon \rightarrow 0$ (where regular black hole solutions have been also found, see \cite{ors15}) and, second, analytical tractability of the corresponding field equations. The latter have been solved and we have found a closed expression for the metric functions. These can be expanded in the region where strong deviations with respect to GR predictions can be found, and their behaviour there splits the analysis into the cases $0<n<1/2$, $1/2<n \leq 1$ and $n>1$. In the cases $1/2<n \leq 1$ we have found that the radial function $r(x)$ reaches a minimum radius $r_c$ at $x=0$, representing the throat of a wormhole which gives a finite structure to the point-like singularity of GR, and smoothly bounces off there, being naturally extended to the $x<0$ region. By studying the behaviour of the metric functions at $r_c$, which is the same for all models belonging to this class, we considered the geodesic equation and showed that, as opposed to the GR counterparts, geodesics can be extended to arbitrarily large values of the affine parameter. Therefore, like in the standard Born-Infeld gravity case, $n=1$, and according to the standard definitions and assumptions employed in the singulary theorems, these solutions represent non-singular spacetimes despite the generic divergence of the curvature scalars at the wormhole throat.

Together with related studies in higher dimensions \cite{or3} or $f(R)$ models recently considered in the literature \cite{fr}, this analysis suggests that wormholes might be quite a generic feature of many Palatini theories of gravity, a fact which could offer new insights for the understanding and avoidance of spacetime singularities in the context of modified theories of gravity. It is worth pointing out that such wormholes are supported by matter sources that satisfy the energy conditions. Moreover, they do not arise as a result of any of the so-called \emph{copy-and-paste} procedures, by which two asymptotically flat spacetimes are matched at a given junction interface (like, for instance, in a thin-shell \cite{TS}) to form a geodesically complete spacetime (see e.g. \cite{Ansoldi} for several spacetimes constructed this way). This is indeed the standard procedure employed to construct wormholes in the context of GR \cite{WHsL}, by which a given wormhole geometry is given first and then the Einstein equations are driven back in order to obtain the matter sources supporting that geometry. As opposed to that case, the wormhole geometries found here flow directly from the resolution of the field equations corresponding to well defined actions. In this sense, the Palatini formulation is also essential to having second-order field equations that can be analytically worked out.

To conclude, let us point out that Born-Infeld gravity and its functional extensions constructed here have been recently shown to be particular cases of a larger family of gravitational actions built as functions of the five polynomial invariants, $e_i(|\hat\Omega|)$, that can be constructed out of the matrix $|\hat\Omega|$, namely

\be
S_{genBI}= \lambda^4 \int d^4x \sqrt{-g} \sum_{n=0}^4 \beta_n e_n(|\hat\Omega|) \ ,
\en
[see Ref.\cite{bho} for definitions and details] where the case just studied in this work corresponds to the polynomials $e_0$ (the identity) and $e_4$ (the determinant). It would be nice to investigate similar electrovacuum scenarios in these theories both to enlarge our knowledge of wormhole physics and to investigate in more detail the avoidance of spacetime singularities in Born-Infeld-type and Palatini theories of gravity. Work along these lines is currently underway.

\section*{Acknowledgments}

C. B. acknowledges support from NSFC (grants 11305038 and U1531117), the Thousand Young Talents Program, and the Alexander von Humboldt Foundation. D.R.-G. is funded by the Funda\c{c}\~ao para a Ci\^encia e a Tecnologia (FCT, Portugal) postdoctoral fellowship No.~SFRH/BPD/102958/2014 and the FCT research grant UID/FIS/04434/2013. This article is based upon work from COST Action CA15117, supported by COST (European Cooperation in Science and Technology).

\end{document}